\documentclass[conference]{IEEEtran}
\usepackage{cite}
\usepackage{graphicx}
\ifCLASSINFOpdf
\else
\fi
\usepackage{algorithm}
\usepackage{algorithmic}
\makeatletter
\renewcommand{\ALG@name}{Motion Control}
\makeatother
\usepackage{url}
\begin{document}
%\title{L\'evy walks having home and scale-free property of human encounters}
%\title{L\'evy walks and scale-free property of human encounters}
\title{Homesick L\'evy walk: A mobility model having Ichi-go Ichi-e and 
scale-free properties of human encounters}

%-----

\author{\IEEEauthorblockN{Akihiro Fujihara$^{1}$ and Hiroyoshi Miwa$^{2}$}
\IEEEauthorblockA{
$^{1}$ Department of Management Information Science, Fukui University of Technology,\\
3-6-1 Gakuen Fukui Fukui 910-8505, JAPAN\\
Email: fujihara@fukui-ut.ac.jp}
\IEEEauthorblockA{
$^{2}$ Graduate School of Science and Technology, Kwansei Gakuin Univ.\\ 
2-1 Gakuen Sanda Hyogo 669-1337, JAPAN\\
Email: miwa@kwansei.ac.jp}
}

%-----


%Telephone: (800) 555--1212, Fax: (888) 555--1212}


\maketitle

\begin{abstract}
%\boldmath
%The abstract goes here.
% The fundamental statistical properties of serendipitous human 
%encounters in daily life are not yet well understood. 
% We conducted experiments to collect long-term data on human 
%contact using short-range wireless communication mobile devices,
%which many people frequently carry.
% By periodically scanning nearby devices, we obtained a reasonably 
%well-sampled data set.
% Our analysis of the data suggested that the majority of human 
%encounters occur once-in-a-experimental-period. 
% We also show that the remaining more frequent encounters obey 
%a power-law type fat-tailed distribution that varies widely. 
% To theoretically explore the origin of these properties, we 
%introduced as a mobility model, the homesick L\'evy walk, 
%a minimal stochastic model of human mobility that traces whether 
%the walker stochastically selects moving long distances as well 
%as L\'evy walk or returning back home.
% Using this model, numerical simulations, and simple mean-field 
%theory, we offer a theoretical explanation for the statistical 
%properties of human contact. 
% We think our mobility model is useful for performance evaluation 
%of routing protocols in delay-tolerant networks since some 
%protocols selects routing paths with frequent encounters in 
%utility-based routing protocols, such as PRoPHET and MAXPROP.
 In recent years, mobility models have been reconsidered based on findings 
by analyzing some big datasets collected by GPS sensors, cellphone call records, 
and Geotagging. 
 To understand the fundamental statistical properties of the frequency of 
serendipitous human encounters, we conducted experiments to collect long-term 
data on human contact using short-range wireless communication devices which 
many people frequently carry in daily life. 
 By analyzing the data we showed that the majority of human encounters
occur once-in-an-experimental-period: they are Ichi-go Ichi-e.
 We also found that the remaining more frequent encounters obey a power-law
distribution: they are scale-free. 
 To theoretically find the origin of these properties, we introduced as a 
minimal human mobility model, Homesick L\'evy walk, where the walker 
stochastically selects moving long distances as well as L\'evy walk or 
returning back home.
 Using numerical simulations and a simple mean-field theory, we offer a 
theoretical explanation for the properties to validate the mobility model. 
 The proposed model is helpful for evaluating long-term performance of routing 
protocols in delay tolerant networks and mobile opportunistic networks better 
since some utility-based protocols select nodes with frequent encounters for 
message transfer.
\end{abstract}



% \fi
\IEEEpeerreviewmaketitle

\section{\label{sec:intro}Introduction}
 In recent years, many kinds of human-carried mobile devices, such as smartphones 
and tablets, that enable many high-tech sensors and wireless communications 
have been increasingly pervasive throughout the world. 
 Because most of the people in the world usually carry these devices in their lives, 
their activity logs, such as places where they visit and persons who they are 
connected with, can be easily recorded using GPS, cellphone call, and Geotagging. 
 We are living in the era of Big data: by analyzing collections of data on people's 
activity, big companies can take advantage of success in their businesses. 
 Also, academic researchers can investigate human activities and social behaviours 
in more details than ever.
 For example, some recent studies based on the analysis of Big data of human mobility 
patterns have revealed that human behavior is easily predictable because human 
mobility is biased in general \cite{SQBB2010,MHVB2013}. 
 This result shows that human mobility is far from random, but is ordered. 
 
 In parallel with the understanding of human mobility patterns, recently, many 
researchers become actively engaged in studies on mobility models, and many mobility 
models have been proposed \cite{santi2012,roy2011}. 
 Traditional mobility models, such as Random Walk (RW), (truncated) L\'evy Walk (LW), 
Random WayPoint (RWP), are simple and basic: they can easily be used for numerical 
simulations in general purposes, but they are far from real human mobility patterns. 
 Therefore, newly proposed models have become more realistic and complicated : They 
includes more parameters to explain many statistical properties on human mobility 
patterns and social effects, thus they resultingly consume more memory space as 
simulation time progresses, such as SLAW \cite{LHKRC2012}. 
 The more a model explains, however, the harder it gets to use for the simulations 
with a large number of walkers in general. 

 These mobility models are often used for evaluating performance of 
routing protocols in Delay Tolerant Networks (DTNs) \cite{VZS2012} and Mobile 
Opportunistic Networks (MON) \cite{denko2011,woungang2013}. 
 To evaluate the performance in systems with a large number of mobile nodes, we 
need to select a balanced mobility model properly. 
 Because the above routing protocols are often contact-based ones, the model 
doesn't necessarily preserve real mobility patterns, but it must have real 
contact patterns. 
 In the context of information communication networks, many researchers frequently 
mention statistical properties on inter-contact time (or inter-meeting time) to 
select the model. 
 But, we would argue in this paper that statistical properties on 
\textit{contact frequency} is also an important factor to properly select the 
mobility model. 

 By the way, there is a famous Japanese proverb closely related 
with the frequency of human contacts which is called \textit{Ichi-go Ichi-e}. 
 This proverb is literally translated as ``One chance in a lifetime'' 
or more specifically as
``Treasure every encounter, for it will never recur.''
 This phrase is closely associated with the history of the tea 
ceremony of Japan.
 \textit{Sen no riky\=u} (1522-1591), the famous tea master during 
the age of the provincial wars, originally taught this proverb to 
his pupils in the spirit of good service.
 Later, at the end of the Edo Period, \textit{Ii Naosuke} (1815-1860),
an accomplished practitioner of the Japanese tea ceremony,
rediscovered and reconsidered this lesson as it is known today.

 We know from experience that to meet with someone (or something) is
sometimes very precious, and we might wonder about how often we have 
once-in-a-lifetime meetings in our daily lives or whether there is 
any statistical law that governs our meetings with people (or things).
 To the best of our knowledge, no scientific study has answered 
these questions because of the difficulty of collecting long-term 
data concerning human contact.
 However, given the recent advances in mobile wireless communication 
technologies, studies on the statistical physics of serendipitous 
human encounters can now be undertaken.

 Our research group collected data concerning daily human contact 
using Bluetooth and Wi-Fi wireless communication technologies.
 Today, billions of electronic devices equipped with Bluetooth and 
Wi-Fi are used throughout the world. 
 Most of these devices are light and mobile, including mobile PCs 
and phones, PDAs, tablets, and portable game machines. 
 Therefore, they tend to be carried at almost all times.
 In addition, the communication range of Bluetooth and Wi-Fi is 
usually on the order of several meters, which is nearly the same 
as the range that humans can see when observing those around them.
 Thus, by scanning and logging nearby Bluetooth and Wi-Fi devices, 
we can collect reasonably well-sampled data concerning human 
encounters. 
 In this case, here, \textit{encounter} should be defined as a state 
that other humans having the device happen to come close within 
several dozen of meters at a maximum distance of its communicable range. 
 In our experiment, we used PDAs and smartphones to continuously scan 
once every twenty seconds and to record pairs of time stamps and MAC addresses 
of detected devices, which indicated when a participant of the 
experiment encountered other people. 
 A sample data logging software that our group uses is available as an 
open-access application for Android OS \cite{pokutuna}.
 To calculate the contact frequency, we need to give a threshold value $\theta_m$ 
where two consecutive device detections whose interval time is less than the 
threshold value are within the same encounter. In this research, we give 
$\theta_m = 1 [\textrm{hour}]$ as a choice, but we also have checked that 
varying the choice of the threshold from some minutes to some hours is not 
sensitive for the whole contact frequency. 
 After conducting our experiment, we obtained \textit{Long data} (rather 
than a recent buzzword \textit{Big data}) whose experimental period is 
between a minimum of several months and a maximum of more than two years. 
(See Table \ref{table:bt_k_val-ichigoichie_ratio} and 
\ref{table:wifi_k_val-ichigoichie_ratio}).
 A dozen of people (university teachers, students, and company workers) 
participated and in total more than 50,000 different devices are 
detected in the experiment. 

 In this paper, by analyzing the collected experimental data, we exhibit 
two basic statistical properties of human contact frequency in the 
long data of human serendipitous encounters: 
(1) the property of Ichi-go Ichi-e, under which most human encounters 
occur once-in-a-experimental-period, and (2) the scale-free property of 
the remaining more frequent meetings. 
 We can find that these properties clearly emerge for each individual 
when analyzing the Long Data. Inversely, it is difficult to observe these properties 
by analyzing short-term data even if they are categorized into the Big Data. 
 To theoretically explain the origin of these statistical properties 
observed in the experiment, we furthermore propose ``Homesick L\'evy Walk'' 
as a simple mobility model. 
 This is a minimal stochastic model of human mobility that traces
whether the walker stochastically selects moving long distances as well 
as the L\'evy walk \cite{bH2000}, or returning back home as a minimum social 
effect. 
 In order to validate the mobility model, we offer a theoretical 
explanation for the properties of human contact using numerical simulations 
and a simple mean-field theory, which is the main contribution of the 
paper. 

 The rest of the paper is organized as follows. 
 In Section~\ref{sec:experiment}, we show the experimental results of 
analyzing collected data with Bluetooth and Wi-Fi opportunistic 
communications.
 In Section~\ref{sec:hlw}, we propose a minimal stochastic model of human 
mobility patterns as a way to simultaneously explain both the two basic
statistical properties.
 In Section~\ref{sec:simulation}, we perform numerical simulations to
show that the minimal model can explain these properties.
 Section~\ref{sec:mft} demonstrates a simple mean-field theory to explain 
the emergence of these properties. 
 We summarize our work, discuss future directions, and comment on an 
application of the model to performance evaluation of routing protocols 
in Delay Tolerant Networks in Sections~\ref{sec:conclusion} and 
\ref{sec:discussions}.

\section{\label{sec:experiment}Experimental results} 
\begin{figure}[tb]
%  \vspace{-25mm}
  \begin{center}
    \includegraphics[width=90mm]{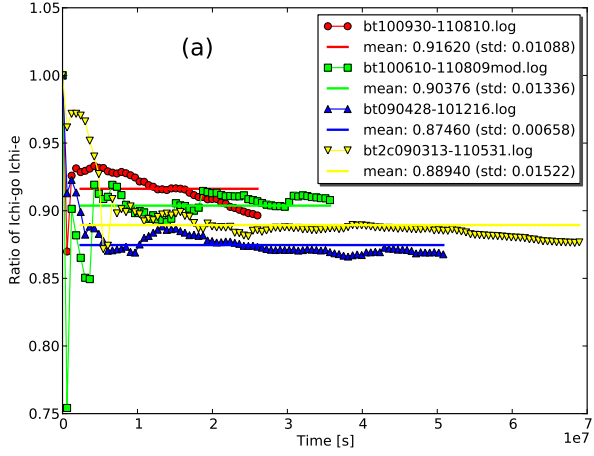}
    \includegraphics[width=90mm]{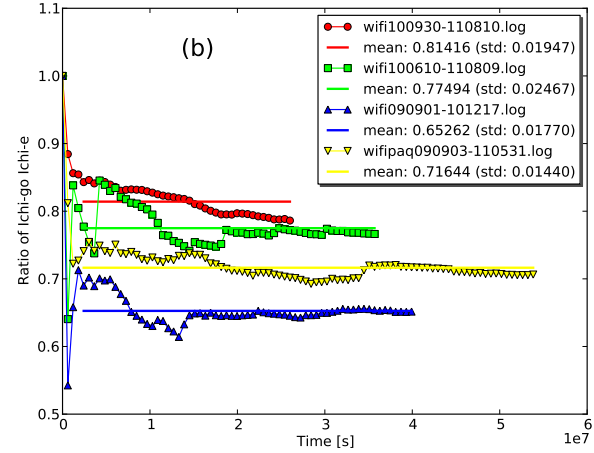}
  \end{center}
%  \vspace{25mm}
  \caption{Time variations in the ratio $R(t)$ for individual 
  participants using (a) Bluetooth and (b) Wi-Fi (line with points)
  and their time-averaged ratios $\langle R \rangle_{t}$ over the 
  experimental time (solid line). In the legends, those of four 
  participants (E, C, B, and A in Tables I and II) are described.}
  \label{fig:ratio_ichigoichie}
\end{figure}

 First, we define $R(t)$ as the ratio of one-time meetings to all 
meetings until time $t$. The ratio of Ichi-go Ichi-e 
(once-in-a-lifetime meetings) to all encounters can also be denoted 
as $R(t)$ at $t=T$ where $T$ is the end of life. 
 Some typical time variations in $R(t)$ during the experiment are 
shown in Fig.~\ref{fig:ratio_ichigoichie}.
 Although the initial patterns of $R(t)$ vary strongly from one 
individual to the next, all of them stabilize as time progresses.
 Therefore, we may roughly assume that each of the time variations 
in $R(t)$ converges to some fixed point around the ratio $R(T)$.
 Under this assumption, we may consider the time-averaged ratio 
over the experimental period denoted as $\langle R \rangle_{t}$ 
to be approximately equivalent to the ratio $R(T)$.
 We present the time-averaged ratio using Bluetooth, 
$\langle R_{bt} \rangle_{t}$, and Wi-Fi, 
$\langle R_{wf} \rangle_{t}$,
in Tables \ref{table:bt_k_val-ichigoichie_ratio} and
\ref{table:wifi_k_val-ichigoichie_ratio}, respectively.
 Averaged over all participants, the percentages of Ichi-go Ichi-e 
meetings using Bluetooth and Wi-Fi are approximately 80-90\%, meaning 
that the majority of human encounters occur once-in-an-experimental-period. 

\begin{figure}[tb]
%  \vspace{-25mm}
  \begin{center}
    \includegraphics[width=90mm]{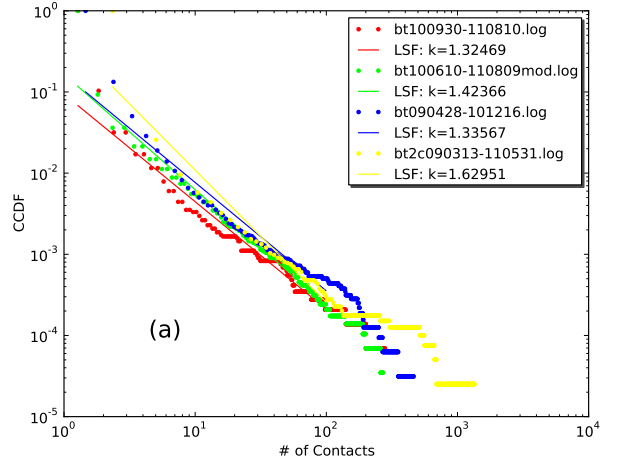}
    \includegraphics[width=90mm]{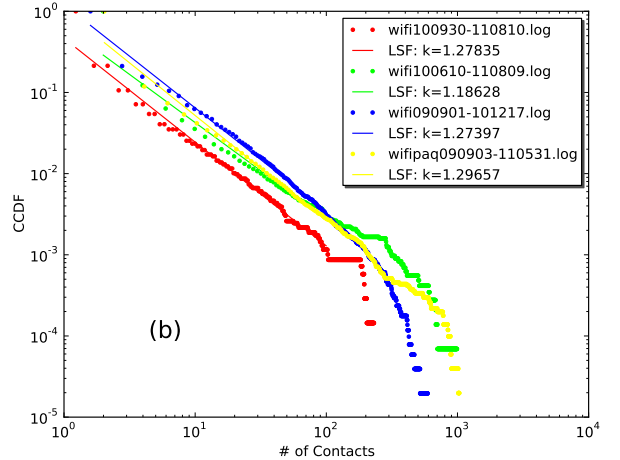}
  \end{center}
%  \vspace{25mm}
  \caption{CCDF for human contact frequency $\bar{F}(x)$ using 
  (a) Bluetooth and (b) Wi-Fi (dots).
  Least squares fitting (LSF) of each CCDF (solid line).
  In the legends, those of four participants (E, C, B, and A in Tables I and II) 
  are described.}
  \label{fig:contact_frequency_dist}
\end{figure}

 We also considered the complementary cumulative distribution
function (CCDF) for human contact frequency. 
 As illustrated in Fig.~\ref{fig:contact_frequency_dist}, the CCDF 
clearly follows a power-law distribution, 
\begin{equation}
F(X \ge x) \equiv \bar{F}(x) \sim  x^{-k},
\end{equation}
where $k$ is a scaling exponent.
 Note that this power law is satisfied only for the remaining 
10-20\% of encounters, those that occur more than once for each 
individual.
 In other words, a large gap exists between $x=1$ and $x \ge 2$ 
in $\bar{F}(x)$.
 The estimated scaling exponents of Bluetooth, $k_{bt}$, and 
Wi-Fi, $k_{wf}$, which were determined using the experimental 
data, are summarized 
in Tables \ref{table:bt_k_val-ichigoichie_ratio} and 
\ref{table:wifi_k_val-ichigoichie_ratio}, respectively.
\begin{table}[t]% Table 1
\caption{The time-averaged ratio of Ichi-go Ichi-e 
$\langle R_{bf} \rangle_{t}$ and the scaling exponent $k_{bf}$ 
for human contact frequency obtained from the Bluetooth data 
for the ten participants.}
\label{table:bt_k_val-ichigoichie_ratio}
\begin{center}
\begin{tabular}{|c|c||c|c|}
\hline
Participant ID & Experimental period & $\langle R_{bt} \rangle_{t}$ & $k_{bt}$ \\
\hline
A  & 2009/03/13-2011/05/31 & $0.89\pm0.02$ & 1.63 \\
B  & 2009/04/28-2010/12/16 & $0.87\pm0.01$ & 1.34 \\
C  & 2010/06/10-2011/08/09 & $0.90\pm0.01$ & 1.42 \\
D  & 2010/09/01-2011/08/08 & $0.87\pm0.01$ & 1.25 \\
E  & 2010/09/30-2011/08/10 & $0.92\pm0.01$ & 1.32 \\
F  & 2010/10/18-2011/02/22 & $0.93\pm0.01$ & 1.24 \\
G  & 2010/10/19-2011/03/09 & $0.93\pm0.01$ & 1.29 \\
H  & 2010/10/21-2011/01/27 & $0.89\pm0.01$ & 1.39 \\
I  & 2010/10/21-2011/02/03 & $0.87\pm0.02$ & 1.17 \\
J  & 2010/11/10-2011/03/31 & $0.84\pm0.03$ & 2.23 \\
\hline
\end{tabular}%
\end{center}
\end{table}
\begin{table}[t]% Table 2
\caption{The time-averaged Ichi-go Ichi-e ratio
$\langle R_{wf} \rangle_{t}$ and the scaling exponent $k_{wf}$ 
for human contact frequency obtained from the Wi-Fi data for
the ten participants.}
\label{table:wifi_k_val-ichigoichie_ratio}
\begin{center}
\begin{tabular}{|c|c||c|c|}
\hline
Participant ID & Experimental period & $\langle R_{wf} \rangle_{t}$ & $k_{wf}$ \\
\hline
A  & 2009/09/03-2011/05/31 & $0.72\pm0.01$ & 1.30 \\
B  & 2009/09/01-2010/12/17 & $0.65\pm0.02$ & 1.27 \\
C  & 2010/06/10-2011/08/09 & $0.77\pm0.02$ & 1.19 \\
D  & 2010/09/01-2011/08/08 & $0.87\pm0.01$ & 1.25 \\
E  & 2010/09/30-2011/08/10 & $0.81\pm0.02$ & 1.28 \\
F  & 2010/10/18-2011/02/22 & $0.80\pm0.03$ & 0.94 \\
G  & 2010/10/19-2011/03/09 & $0.82\pm0.01$ & 0.90 \\
H  & 2010/10/21-2011/01/27 & $0.73\pm0.02$ & 1.40 \\
I  & 2010/10/21-2011/02/03 & $0.73\pm0.01$ & 1.42 \\
J  & 2010/11/10-2011/03/31 & $0.63\pm0.01$ & 1.42 \\
\hline
\end{tabular}%
\end{center}
\end{table}
 For almost all the participants, the estimated scaling exponents 
are lower than two. Because of this, the variance of the CCDF tends 
to diverge during a long experimental period. 
 This divergence indicates that the number of encounters that the 
participants meet with people in their experimental periods has no 
characteristic scale. 
 In other words, meetings inherently exhibit extreme inequality, 
making it difficult to predict how many opportunities are left to 
encounter someone (or something).

\section{\label{sec:hlw}Homesick L\'evy Walk}
 Next, we propose a minimal stochastic model of human mobility 
patterns as a way to simultaneously explain both the Ichi-go Ichi-e 
and scale-free properties of human encounters.
 A number of researchers have recently reported that human mobility 
traces statistically exhibit the L\'evy walk 
(LW) \cite{BHG2006,RSHLC2008,GHB2008,Barabasi2010}. 
 A L\'evy walker in $d$-dimensional space determines his or her 
destination; the travel distance from the present location $l$ is 
governed by an 
independently and identically distributed power-law distribution, 
\begin{equation}
p(l) \sim l^{-(1+\beta)}, \label{eq:levy_walk}
\end{equation}
where $0 < \beta \le 2$ is the L\'evy index. 
 The direction from the present location to the next destination 
location is usually determined by the uniform distribution. 
 This feature has been found to be common to the mobility of 
humans and animals in two-dimensional space
\cite{VBHLRS1999,EPWFMABLRSV2007,SSHHBPJABHMMRSWWWM2008,VLRS2011}. 

\begin{figure}[tb]
%  \vspace{-25mm}
  \begin{center}
    \includegraphics[width=80mm]{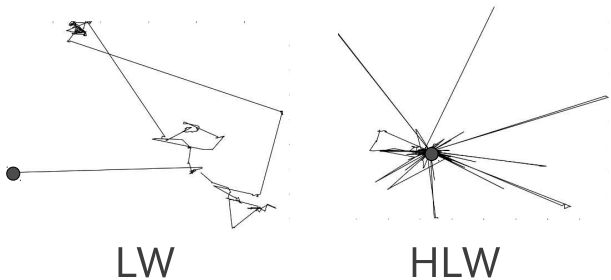}
  \end{center}
%  \vspace{25mm}
  \caption{Typical sample traces for L\'evy walk (LW) and homesick 
  L\'evy walk (HLW). 
   The initial position of each walker is indicated by the grey 
  circle, which is assumed to be the hub of activity (or home) in 
  the HLW scenario.}
  \label{fig:lw_vs_hlw}
\end{figure}

 However, the scale-free property of walk lengths is insufficient 
to explain the statistical properties of human contact frequency. 
 We numerically confirmed that the CCDF of contact frequency for
L\'evy walkers in bounded two-dimensional space generally decays 
exponentially at the tail.
 Let us consider what is lacking in the ordinary L\'evy walk scenario.
 A L\'evy walker easily travels long distances, but has difficulty 
returning to his or her original position (refer to 
Fig.~\ref{fig:lw_vs_hlw}).
 In real life, however, each participant typically frequents 
his or her own hub of social activity (or his or her own home). 
This reality strongly determines most of the topology of human 
mobility traces.
 Taking into consideration the role of the hub, we propose an 
extended version of the L\'evy walk named the ``Homesick L\'evy 
Walk (HLW).'' In this model, after arriving at the destination 
determined by the power-law walk length in Eq.~(\ref{eq:levy_walk}), 
there exists a certain fixed probability $\alpha$ that the L\'evy 
walker will become \textit{homesick} and return home; otherwise, 
according to probability $(1-\alpha)$, the walker will determine his 
or her next destination using Eq.~(\ref{eq:levy_walk}) and continue 
travelling. By definition, HLW with a homesick probability of 
$\alpha=0$ reduces to LW. 
 For the sake of convenience, we define the initial position 
of HLW as home.
 The detailed procedure to move two-dimensional HLW is shown 
in Motion Control \ref{alg:hlw}.
 The difference between the sample traces associated with the 
(simple) L\'evy walk and the homesick L\'evy walk is illustrated 
in Fig. \ref{fig:lw_vs_hlw}.

 In the name of this model, we use the word \textit{walk} but not 
\textit{flight}. In general, \textit{walk} means that a walker 
moves with a finite velocity to the destination, but \textit{flight} 
means that a walker jump instantly to the destination, which is the 
difference between these words. Because we also consider serendipitous 
encounters on the way to the destination, we introduce homesick L\'evy 
\textit{walk} here. But, Homesick L\'evy Flight (HLF) can be defined 
by a similar way with changing from walk to flight in the model. 

\begin{algorithm}[h!]
    \caption{: (Two-dimensional) Homesick L\'evy Walk}\label{alg:hlw}
    \begin{algorithmic}
    \REQUIRE Initial position: Hub of activity (or home) $x_{home}$  (The current position $x_{now} = x_{home}$), Initial state: State == Stop; the homesick probability $\alpha$; the scaling exponent of the CCDF of contact frequency $\beta$
    \WHILE{TRUE (The walker is alive.)}
        \IF{ State == Stop }
            \STATE State == Move
            \IF{ Probability: $\alpha$ }
                \STATE Destination == Hub of activity ($x_{dest} = x_{home}$)
            \ELSE
                \STATE New destination $x_{dest}$ = $x_{now}$ + $l$ $\times$ ($\cos{\theta}$, $\sin{\theta}$), where $l$ is determined by Eq.~(\ref{eq:levy_walk}) and the angle $\theta$ 
                \STATE is given by the uniform distribution on $[0,2\pi)$. 
            \ENDIF
        \ENDIF
        \STATE Start moving to the destination $x_{now} \gets x_{now} + \Delta l \times$ ($\cos{\theta}$, $\sin{\theta}$), where $\Delta l$ is the distance that the walker moves in a single step.
        \IF{ $x_{now}$ == $x_{home}$ or $x_{dest}$ }
            \STATE State == Stop
            \STATE (The walker may wait for a while.)
        \ENDIF
    \ENDWHILE
    \end{algorithmic}
\end{algorithm}

\section{\label{sec:simulation}Numerical Simulations}
 We performed numerical simulations for $N$ homesick L\'evy walkers 
in a bounded two-dimensional space. Initially, $N=1000$ walkers were 
uniformly distributed within a $1726 [\textrm{m}] \times 1726 [\textrm{m}]$ 
squared region. 
 The number of the walkers $N$ does not change in the simulations. 
 The density of the walkers was determined based on the average 
population density of Japan, approximately $336 [\textrm{km}^{-2}]$, 
because all of the study participants live mainly in Japan. 
 The distance between the present position and the next destination 
described using the polar coordinate $l = (r \cos \theta, r \sin \theta)$
is randomly generated from the probability distribution function, 
$p(l) \equiv p(r) p(\theta)$, where
\begin{equation}
p(r) = \frac{\beta r_{m}^{\beta}}{r^{2+\beta}} \hspace{3mm} (r > r_{m}),
\hspace{5mm}
p(\theta) = \frac{1}{2\pi} \hspace{3mm} (0 \le \theta < 2 \pi),
\label{eq:2d_levy_walk_r_theta} 
\end{equation}
and the minimum travelling distance $r_{m}=1 [\textrm{m}]$.
 These equations are derived from Eq.~(\ref{eq:levy_walk}) to 
satisfy the normalization condition $\int_{r_{m}}^{\infty} p(l) dl = 1$.
 We assumed that for each time-step, all of the walkers move with a 
constant speed $v=1 [\textrm{m/s}]$, and a small number of walkers meet 
together if they are within a fixed communication radius $c=1 [\textrm{m}]$.
 We also assumed that after a walker arrive at a destination, 
the walker waits a single time-step to determine the next destination.

\begin{figure}[tb]
%  \vspace{-25mm}
  \begin{center}
    \includegraphics[width=90mm]{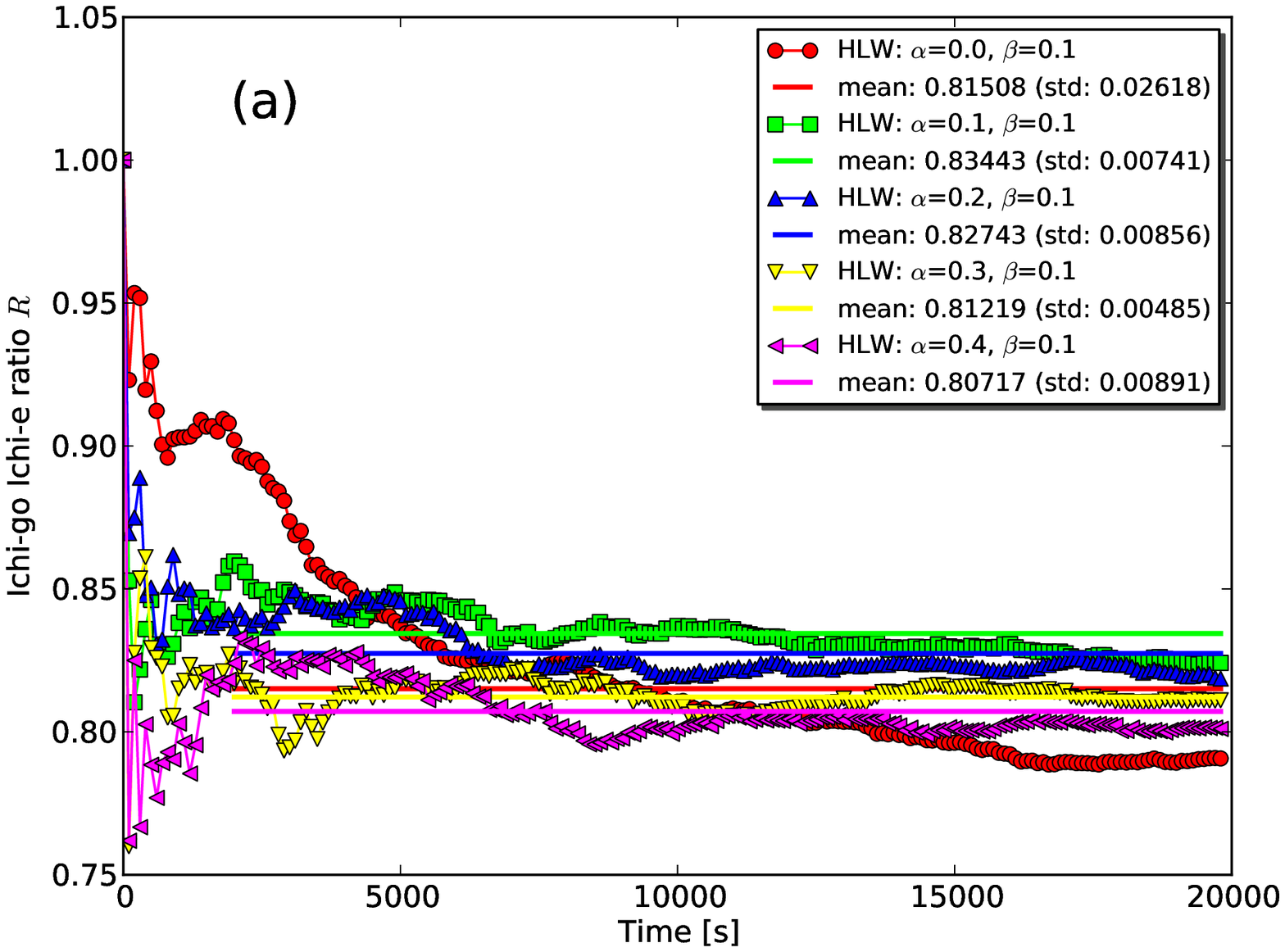}
    \includegraphics[width=90mm]{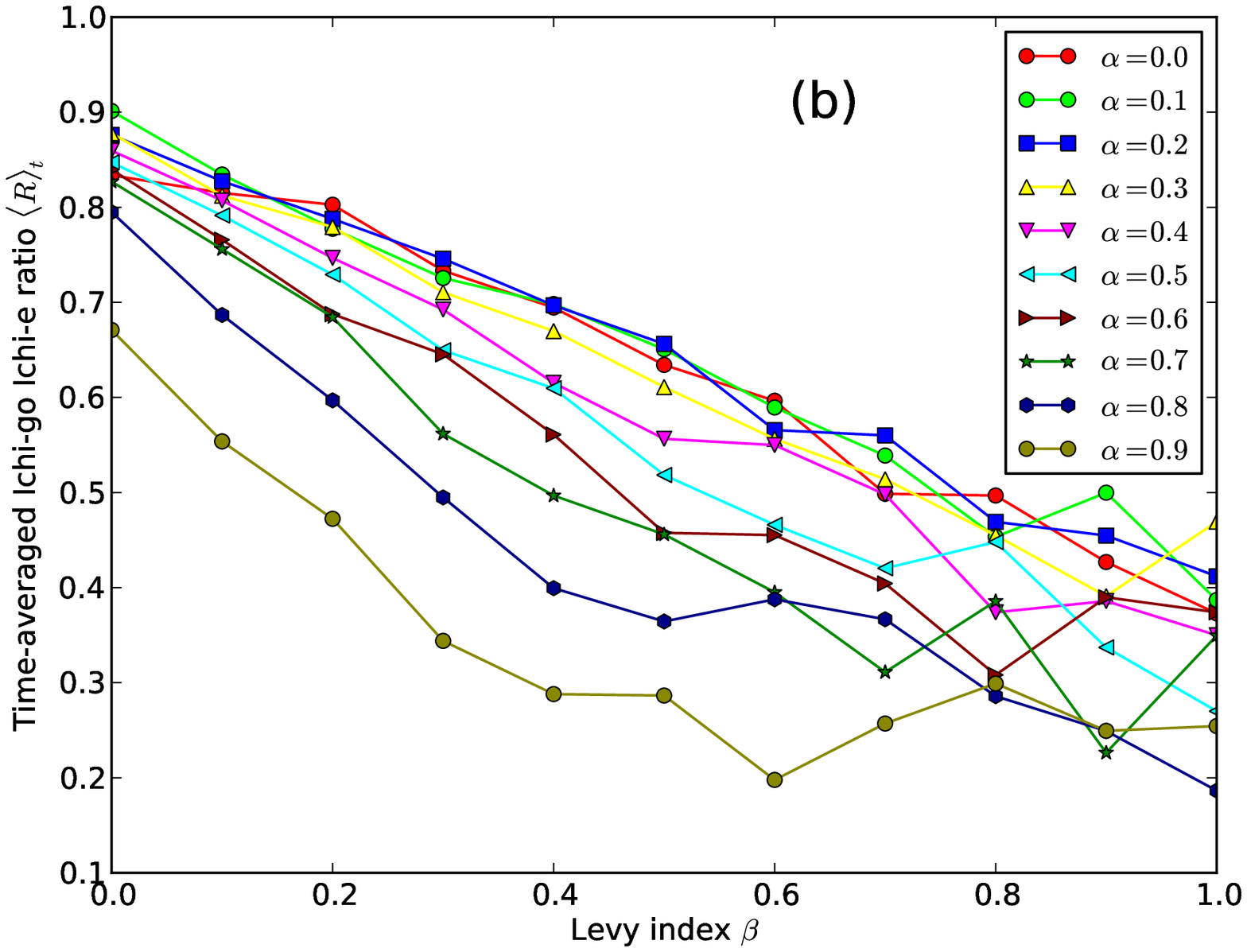}
  \end{center}
%  \vspace{25mm}
  \caption{(a) The typical time evolution of $R(t)$ in the L\'evy 
  walk ($\alpha=0$) and the homesick L\'evy walk ($\alpha>0$)
  determined using numerical simulations for $N=10^3$, 
  $T=2.0 \times 10^4 [\textrm{s}]$, $\alpha=0, 0.1, 0.2, 0.3, 0.4$, and 
  $\beta=0.1$ (lines with dots).
  The time-averaged Ichi-go Ichi-e ratios $\langle R \rangle_{t}$ 
  are also included (solid line).
  (b) Relationship between $\langle R \rangle_{t}$ and $\alpha$ and
  $\beta$ over the ranges $0 \le \alpha < 1$ and 
  $0 \le \beta \le 1$.}
  \label{fig:alpha_beta_vs_R}
\end{figure}

 Next, we considered the effects of the homesick property and 
long-distance travelling on the time-averaged Ichi-go Ichi-e ratio
$\langle R \rangle_{t}$ and the scaling exponent of the CCDF of 
the contact frequency $k$, varying $\alpha$ and $\beta$. 
 The typical time evolution of $R(t)$ is illustrated in 
Fig.~\ref{fig:alpha_beta_vs_R} (a).
 As observed in Fig.~\ref{fig:ratio_ichigoichie}, each $R(t)$
tends to converge as time progresses.
 The time-averaged ratio $\langle R \rangle_{t}$, calculated by 
averaging $R(t)$ over the simulation time, is also plotted for 
the ranges $0 \le \alpha < 1$ and $0 \le \beta \le 1$ in 
Fig.~\ref{fig:alpha_beta_vs_R} (b).
 We numerically confirmed that $\langle R \rangle_{t}$ tends to 
decrease with increases in $\alpha$ and $\beta$.
 Also, we observe that by roughly tuning parameters $\alpha$ and 
$\beta$ to sufficiently small values, we obtain a value of more than
80\% for $\langle R \rangle_{t}$ and the numerical results becomes 
consistent with the experimental ones.

\begin{figure}[tb]
%  \vspace{-25mm}
  \begin{center}
    \includegraphics[width=90mm]{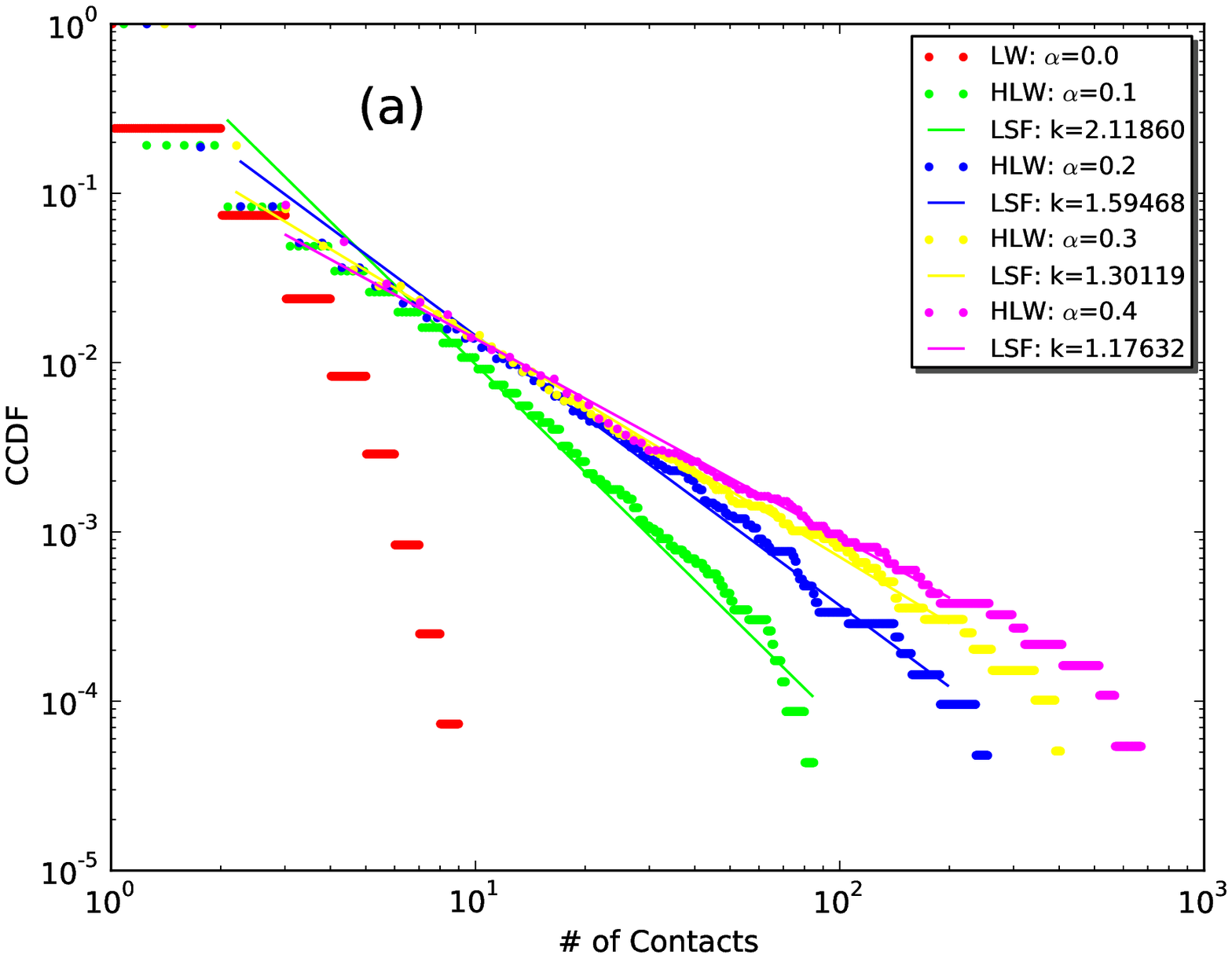}
    \includegraphics[width=90mm]{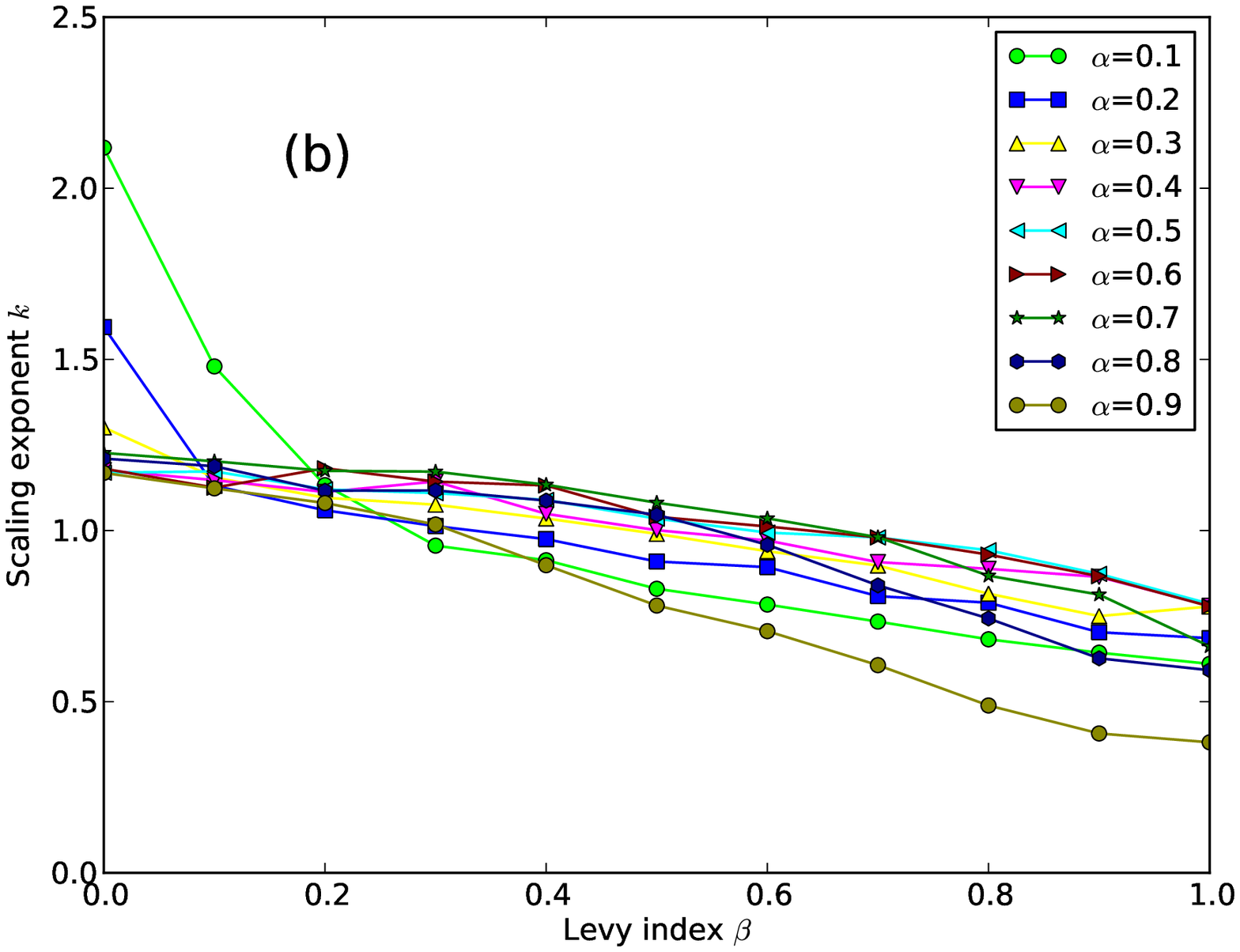}
  \end{center}
%  \vspace{25mm}
  \caption{(a) Typical CCDFs for contact frequency values 
  $\bar{F}(x)$ employing LW and HLW in numerical simulations 
  where $N=10^3$, $T=10^5 [\textrm{s}]$, and 
  $\alpha=0, 0.1, 0.2, 0.3, 0.4$, $\beta=0$ (dots) with least squares 
  fitting (LSF) for the CCDFs (solid line).
  (b) Relationship between the estimated scaling exponent $k$ 
  from the CCDFs for contact frequency and $\alpha$ and $\beta$ for 
  $0 < \alpha < 1$ and $0 \le \beta \le 1$.}
  \label{fig:ccdf_hlf_vs_lf}
\end{figure}

 In Fig. \ref{fig:ccdf_hlf_vs_lf}(a), we also present some typical 
CCDFs for contact frequency based on the numerical results.
 We can clearly see that the CCDF created using HLW for $\alpha>0$ 
obeys a power-law distribution, whereas that created using LW 
($\alpha=0$) decays exponentially at the tail, as we mentioned before.
 This exponential decay for $\alpha=0$ was observed for the entire 
range, $0 \le \beta \le 1$. The results indicate that the 
homesickness component of HLW is essential to the scale-free property 
of contact frequencies.
 We also consider the relationship between the scaling exponent $k$ 
for contact frequency and for $\alpha$ and $\beta$ calculated using 
least-squares fitting. 
 As observed in Fig. \ref{fig:ccdf_hlf_vs_lf}(b), $k$ has a weak 
decreasing trend with increasing $\alpha$ and $\beta$ for $0 < \alpha < 1$
and $0 \le \beta \le 1$.
 The most important point is that the value of $k$ matches the 
experimental values of $k_{bt}$ and $k_{wf}$ in Tables 
\ref{table:bt_k_val-ichigoichie_ratio} 
and \ref{table:wifi_k_val-ichigoichie_ratio} when we keep $\alpha$ ($>0$) 
and $\beta$ small, as shown in Fig. \ref{fig:ccdf_hlf_vs_lf}(b).
 As a result, we numerically demonstrate that the HLW stochastic 
model can explain human contact frequency. 

 It should be emphasized that the contact frequencies of 
Homesick Random Walk (HRW) and 
\begin{figure}[tb]
%  \vspace{-25mm}
  \begin{center}
    \includegraphics[width=85mm]{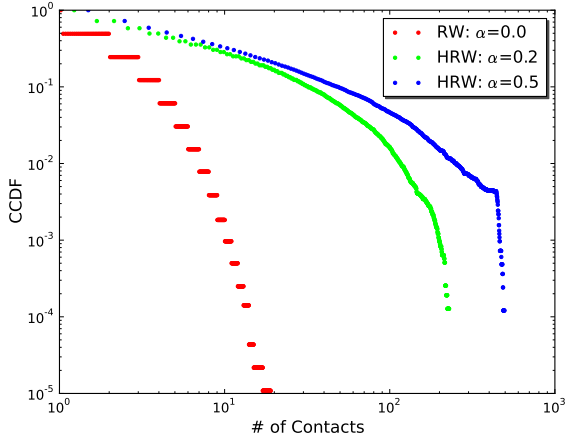}
  \end{center}
%  \vspace{25mm}
  \caption{Typical CCDFs for contact frequency values 
  $\bar{F}(x)$ employing RW and HRW in numerical simulations 
  where $N=10^3$, $T=10^5 [\textrm{s}]$, and 
  $\alpha=0, 0.2, 0.5$ (dots).}
  \label{fig:ccdf_hrw_vs_rw}
\end{figure}
Homesick Random WayPoint (HRWP)
\begin{figure}[tb]
%  \vspace{-25mm}
  \begin{center}
    \includegraphics[width=85mm]{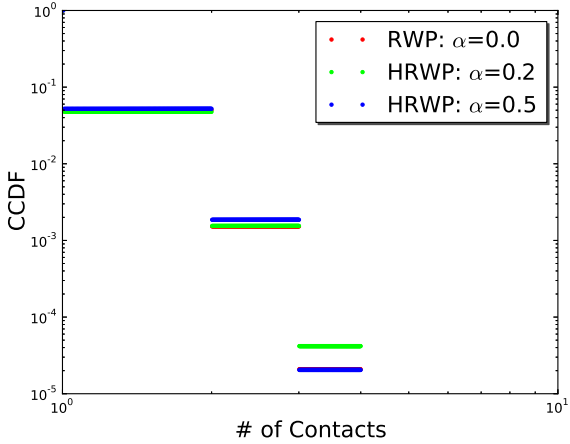}
  \end{center}
%  \vspace{25mm}
  \caption{Typical CCDFs for contact frequency values 
  $\bar{F}(x)$ employing RWP and HRWP in numerical simulations 
  where $N=10^3$, $T=10^5 [\textrm{s}]$, and 
  $\alpha=0, 0.2, 0.5$ (dots).}
  \label{fig:ccdf_hrwp_vs_rwp}
\end{figure}
seem to be different from that of HLW as shown in 
Figs.~\ref{fig:ccdf_hrw_vs_rw} and \ref{fig:ccdf_hrwp_vs_rwp}, 
thus tail's fatness of these distribution is less than that of 
the power-law distribution.

\section{\label{sec:mft}Mean-field Theory}
\begin{figure}[tb]
%  \vspace{-25mm}
  \begin{center}
    \includegraphics[width=65mm]{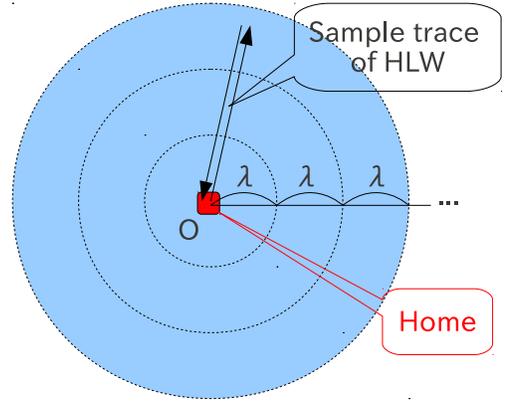}
  \end{center}
%  \vspace{25mm}
  \caption{ Schematic illustration of the concentric zone hypothesis.}
  \label{fig:czh}
\end{figure}

 Finally, we use a simple mean-field theory of HLW to explain the 
emergence of these phenomena. 
 We focus on one walker whose home is fixed at the origin of 
the two-dimensional space, whereas the other walkers are assumed 
to be spatially fixed and uniformly distributed in the space. 
 We also define the mean-free path $\lambda$ as the averaged 
moving distance that one traverses before encountering the next 
walker. 
 Because the walker continues to repeatedly travel around and 
return home, the spatial existence probability of the focused 
walker tends to increase as the distance from home decreases.
 With this viewpoint in mind, we assume that the contact 
frequency for this walker and others depends only on the distance 
from home $r'$ and that the walker meets the same walker at the 
same distance.
 Taking into consideration all of the above definitions and 
assumptions, we envision a scenario in which the walker encounters 
a new walker whenever he walks away from the concentric rings
of width $\lambda$ whose centre is his home. 
 This image of meetings with walkers is used to separate the space 
into many ring-shaped concentric zones in which the walker in
question meets others, as shown in Fig.~\ref{fig:czh}. 
 We call this the \textit{concentric zone hypothesis}.

\begin{figure}[tb]
%  \vspace{-25mm}
  \begin{center}
    \includegraphics[width=90mm]{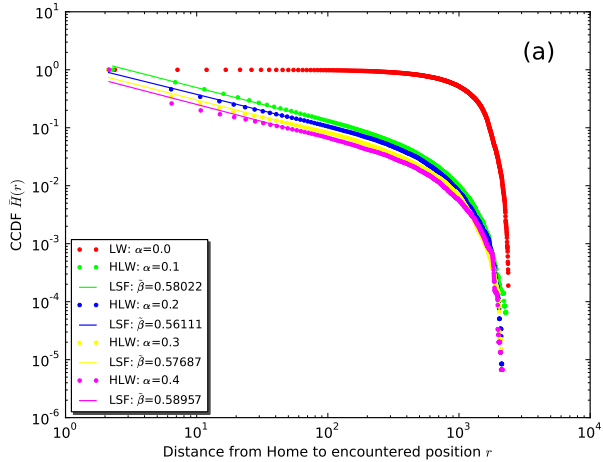}
    \includegraphics[width=90mm]{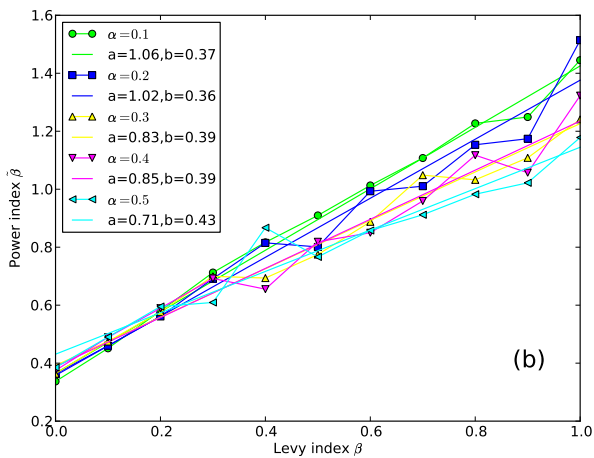}
  \end{center}
%  \vspace{25mm}
  \caption{ (a) A typical spatial existence probability distribution
  $\bar{H}(r' \ge r)$ for HLWs where $N=10^3$, $T=10^5 [\textrm{s}]$, 
  $\alpha=0, 0.1, 0.2, 0.3, 0.4$, and $\beta=0.2$ (dots).
  The power index $\tilde{\beta}$ is also shown using the least-square 
  fits for the distribution (solid line).
  (b) The relationship between the parameters 
  $\alpha=0.1, 0.2, 0.3, 0.4, 0.5$ and $0 \le \beta \le 1$ and 
  the estimated power index $\tilde{\beta}$ (line with dots) and their
  least-square fits (line).}
  \label{fig:ccdf_dfh}
\end{figure}

 In Fig.~\ref{fig:ccdf_dfh}(a), we present numerical results for 
the CCDF of the walker's frequency of contact with others with 
respect to his distance from home and the location at which he
encounters those other individuals, $\bar{H}$.
 We observe that the function $\bar{H}$ generally obeys a power-law 
distribution for $\alpha>0$, especially for the lower range of $r$, 
\begin{equation}
\bar{H}(r' \ge r) \sim  r^{-\tilde{\beta}},
\end{equation}
where $\tilde{\beta}$ is the power index.
 The cut-off of the power law at the tail arises from the effect of 
spatial boundedness. For $\alpha=0$, however, $\bar{H}$ is more 
likely to behave like a uniform distribution than it is to behave like 
a power law.
 Therefore, the spatial existence probability $\bar{H}$ appears to be 
qualitatively different for HLW and LW.

 In Fig.~\ref{fig:ccdf_dfh} (b), we showed the numerical results 
indicating how the power index $\tilde{\beta}$ behaves given changes
in $\alpha$ and $\beta$. We can see that a proportional relationship
between $\tilde{\beta}$ and $\beta$, 
\begin{equation}
\tilde{\beta} \simeq c \beta + d, \label{eq:prop_rel}
\end{equation}
is approximately satisfied for small $\alpha$ and $0 \le \beta \le 1$.
 The proportionality coefficient $c(\alpha)$ in Eq.~(\ref{eq:prop_rel}), 
which seems to have a particular value when $\alpha$ is fixed, tends to 
gradually decrease with increasing $\alpha$.
 These values of $c(\alpha)$ for $\alpha=0.1,0.2,0.3,0.4,0.5$ varies 
around one.

 Applying the power-law distribution of $\bar{H}$ to the concentric 
zone hypothesis allows us to calculate the contact frequency of the 
walker in the $j$-th concentric ring zone, denoted as $x_j$. 
 For this purpose, we use $h(r) \equiv - d \bar{H}(r' \ge r)/dr$ as 
follows:
\begin{eqnarray}
x_j &\approx& \int_{j\lambda}^{(j+1)\lambda} h(r) r dr 
\sim \int_{j\lambda}^{(j+1)\lambda} r^{-(1+\tilde{\beta})} r dr \nonumber \\
    &\simeq & \lambda^2 j/(j\lambda)^{1+\tilde{\beta}} \propto j^{-\tilde{\beta}}. \label{eq:rank_dist}
\end{eqnarray}
 This equation directly indicates that the rank distribution of 
the contact frequencies also obeys a power-law distribution whose 
power index is $\tilde{\beta}$.
 When $\beta$ remains within its small range $0 \le \beta < (1-d)/c$, 
the power index of the rank distribution $\tilde{\beta}$ becomes 
less than one by Eq.~(\ref{eq:prop_rel}). 
 In this case, therefore, the tail of the rank distribution is 
generally so wide that the number of low-ranked walkers is divergent, 
which provides a simple explanation of why the majority of human 
encounters is Ichi-go Ichi-e.

\begin{figure}[tb]
%  \vspace{-25mm}
  \begin{center}
    \includegraphics[width=90mm]{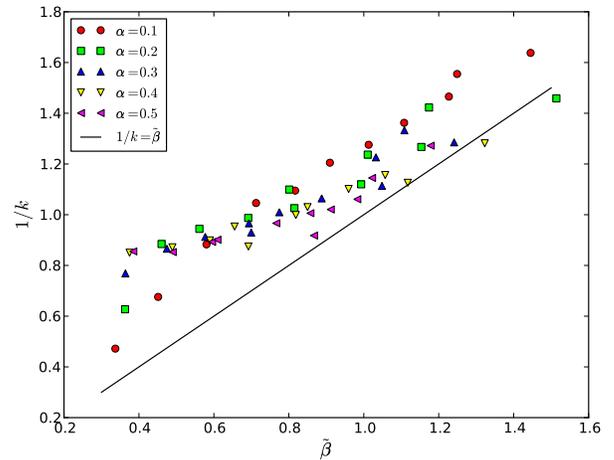}
  \end{center}
%  \vspace{25mm}
  \caption{Relation between $k$ and $\tilde{\beta}$ for 
  $\alpha=0.1,0.2,0.3,0.4,0.5$ and $0 \le \beta \le 1$ (dots) and 
  the exact inverse relationship in Eq.~(\ref{eq:k_relation}) (solid line).}
  \label{fig:inv_rel}
\end{figure}

 We also consider the relationship between the rank distribution
and the CCDF for contact frequency.
 It is well-known that if a rank distribution obeys a 
power law, then its frequency also becomes a power law. Thus, for 
the rank distribution $x_j$, we can obtain the scale-free human 
contact frequency.
 In this case, it is also known that an inverse relationship exists 
between $k$ and $\tilde{\beta}$: 
\begin{equation}
k = 1/\tilde{\beta} \simeq 1/(c \beta + d). \label{eq:k_relation}
\end{equation}
 We checked whether this inverse relation is numerically supported 
in Fig.~\ref{fig:inv_rel}.
 We can see that Eq.~(\ref{eq:k_relation}) roughly explains the trend 
between $k$ and $\tilde{\beta}$. Substituting controlled values of 
$0 \le \beta < (1-d)/c$ into Eq.~(\ref{eq:k_relation}) yields values of 
$k$ ($>1$) that are consistent with the experimental presented 
in Tables \ref{table:bt_k_val-ichigoichie_ratio} 
and \ref{table:wifi_k_val-ichigoichie_ratio}.

\section{\label{sec:conclusion}Conclusion}
 In summary, we investigated the general statistical properties 
of serendipitous human encounters in daily life using portable 
wireless communication devices. We experimentally determined that
we can universally apply the following statistical principles 
to human contact frequencies among individuals: 
\begin{enumerate}
  \item The majority of human encounters occur once during one's 
        experimental period and this feature seems to continue 
        during one's lifetime (the property of Ichi-go Ichi-e)
  \item The remaining, more frequent encounters obey the power-law 
        distribution in terms of contact frequency and its variance
        diverges (the scale-free property)
\end{enumerate}
 Because this long tail principle of human contact frequency is
universal, it is difficult to predict how many times we will have 
additional opportunities to meet with people over long periods 
of time. This further validates the principle of Ichi-go Ichi-e, 
which teaches us to ``treasure every encounter.''

 To facilitate a fundamental theoretical understanding of these 
principles, we introduced a novel stochastic model of human mobility 
traces called the ``homesick L\'evy walk.''  
 We used numerical simulations to demonstrate that this model can 
successfully explain both of the principles.
 Furthermore, using simple mean-field theory, we determined that 
the origin of the principles arises from the following two opposing 
mechanisms that inherently underlie human mobility patterns: 
\begin{enumerate}
  \item Long-distance travelling ($0 \le \beta <(1-d)/c$ in HLW)
  \item Homesickness ($\alpha>0$ in HLW)
\end{enumerate}
 Although we have proposed this model in \cite{FM2011b,FM2013} 
for evaluation our proposed routing method, the origin of the above 
properties is first explained in this paper. 
 Balancing the two mechanisms leads the statistical principles of 
human contact frequency to emerge. Note that according to violating 
the first mechanism, ``Homesick Random Walk (HRW)'' whose walk length 
to the next destination is determined by a distribution with a finite 
variance does not have the scale-free property of contact frequency, 
which we have also checked numerically. 

 It should be noted that we have introduced \textit{home} as a minimum 
social effect. Because homesick L\'evy walkers periodically return to 
their home, they tend to stay longer around their homes. Therefore, 
they meet with each other more frequently as the distance between their 
homes is closer, which naturally includes social relations between 
the walkers.

\section{\label{sec:discussions}Discussions}

 In this paper, we focused on contact frequency for \textit{humans}.
 However, that of \textit{animals} might also obey the same principles 
because the above two mechanisms are common to animals and humans: 
animals usually have nests that are similar to our homes, and they 
also travel to distant feeding sites. 

 For future works, it is also important to consider effects of 
non-uniformity of population density. In the simulations, we used 
the average population density of Japan, but the density of people 
highly varies by the size of city where they live. Effects of the 
density on contact frequency, inter-contact time, and contact 
duration is an interesting task to investigate. 

 Recently, Song \textit{et al.} has proposed ``preferential return''
to explain spatial visitation properties of human mobility patterns 
using their individual mobility model \cite{SKWB2010}. 
 Our homesick L\'evy walk model seems to be similar to their model, 
but there are some differences: Since they consider the 
\textit{visitation} frequency of locations, their model needs multiple 
locations where one can return preferentially by \textit{flight}. 
 Since we consider the \textit{contact} frequency between humans, 
on the other hand, our model does not necessarily assume multiple 
locations, but only one hub location where one can return with a 
fixed probability by \textit{walk}. 
 By experiment, we empirically know that the serendipitous human 
contacts occur on the way to destination more frequently rather than 
duration of visit at destination. 
 Therefore, theoretical results given by their model does not cover 
statistical properties of the majority of human contacts.
 To understand the relation between these models is left for future
work.

 We think our mobility model is useful for performance evaluation 
of routing protocols in Delay Tolerant Networks (DTN) \cite{VZS2012} 
since some protocols selects routing paths with frequent encounters
in utility-based routing protocols, such as PRoPHET \cite{Lindgren2004}, 
MAXPROP \cite{Burgess2006}, and so on. 
 We also have proposed our algorithm for routing in DTN 
and have shown some results regarding the comparison between LW and 
HLW \cite{FM2011b,FM2013}, which indicates that the arrival rate of 
transferred messages tends to be much lower as increasing the 
homesick probability $\alpha$. It is also important to take into 
consideration the effect of Ichi-go Ichi-e since the large number of 
human encounters is rare. Therefore, the majority of human encounters 
usually doesn't contribute to the performance of the utility-based routing, 
but they only consume much memory in vain for memorizing the history of 
encounters with many nodes that will never be encountered again. 
 After understanding the properties of the frequency of human contact well, 
the routing protocols for message transfer could be improved.

\section*{Acknowledgment}
 This work is partially supported by Japan Society for the Promotion 
of Science through Grant-in-Aid for Scientific Research (C) (23500105) 
and Grant-in-Aid for Young Scientists (B) (25870958).



%\bibliographystyle{IEEEtran}

\begin{thebibliography}{99}

% predictable, human mobility
\bibitem{SQBB2010}
  Song, C., \textit{et al.}
  Limits of Predictability in Human Mobility,
  Science, vol. 327, no. 5968, pp.1018-1021 (2010).

\bibitem{MHVB2013}
  Y.-A. de Montjoye, \textit{et al.},
  Unique in the Crowd: The privacy bounds of human mobility,
  Scientific Reports, vol. 3, no. 1376 (2013).

% mobility models
\bibitem{roy2011}
  Roy, R. R.
  \textit{Handbook of Mobile Ad Hoc Networks for Mobility Models}
  (Springer, 2011).


\bibitem{santi2012}
  Santi, P.
  \textit{Mobility Models for Next Generation Wireless Networks: 
  Ad Hoc, Vehicular and Mesh Networks}
  (Wiley, 2012).


% SLAW
\bibitem{LHKRC2012}
  Lee, K. \textit{et al.}
  SLAW: Self-Similar Least-Action Human Walk,
  IEEE/ACM Transactions on Networking, vol. 20, no.2, (2012).


% opportunistic networks
\bibitem{denko2011}
  Denko, M. K.
  \textit{Mobile Opportunistic Networks: Architectures, Protocols and Applications}
  (Auerbach Publications, 2011). 

\bibitem{woungang2013}
  Woungang, I., \textit{et al.}
  \textit{Routing in Opportunistic Networks}
  (Springer, 2013).

% android 
\bibitem{pokutuna}
  \url{http://tech.pokutuna.com/android-wireless-device-logger/}

\bibitem{bH2000}
    ben-Avraham, D. \& Havlin, S.
    \textit{Diffusion and Reactions in Fractals and Disordered Systems}
    (Cambridge University Press, 2000).

\bibitem{Barabasi2010}
    Barab\'asi, A.-L.
    \textit{Bursts: The Hidden Pattern Behind Everything We Do}
    (Dutton Adult, 2010).

\bibitem{BHG2006}
    Brockmann, D. \textit{et al.}
    The scaling laws of human travel.
    \textit{Nature} \textbf{439} 462-465 (2006).

\bibitem{GHB2008}
    Gonz\'alez, M. C. \textit{et al.}
    Understanding individual human mobility patterns.
    \textit{Nature} \textbf{453}, 779-782 (2008).

\bibitem{RSHLC2008}
    Rhee, I. \textit{et al.}
    On the Levy-walk Nature of Human Mobility: Do Humans Walk like Monkey?
    The 27th IEEE International Conference on Computer Communications 
    (IEEE INFOCOM 2008), 924-932 (2008).

\bibitem{VBHLRS1999}
    Viswanathan, G. M. \textit{et al.}
    Optimizing the success of random searches. 
    \textit{Nature} \textbf{401}, 911–914 (1999).

\bibitem{EPWFMABLRSV2007}
    Edwards, A. M. \textit{et al.}
    Revisiting L\'evy flight search patterns of wandering albatrosses, 
    bumblebees and deer.
    \textit{Nature} \textbf{449}, 1044-1048 (2007).

\bibitem{SSHHBPJABHMMRSWWWM2008}
    David W. Sims \textit{et al.}
    Scaling laws of marine predator search behaviour.
    \textit{Nature} \textbf{451}, 1098-1102 (2008).

\bibitem{VLRS2011}
    Viswanathan, G. M.  \textit{et al.}
    \textit{The Physics of Foraging: An Introduction to Random Searches 
    and Biological Encounters}
    (Cambridge University Press, 2011).

% preferential return
\bibitem{SKWB2010}
    C. Song \textit{et al.},
    Modelling the scaling properties of human mobility,
    Nature Physics (Advanced Online Publications) 7, 713 (2010).

% DTN
\bibitem{VZS2012}
  A.~Vasilakos, Y.~Zhang, and T.~V. Spyropoulos,
  \emph{Delay Tolerant Networks: Protocols and Applications},
  Wireless Networks and Mobile Communications Series,
  CRC Press, 2012.

%PRoPHET
\bibitem{Lindgren2004}
  A. Lindgren, A. Doria, and O. Schel\'en, 
  ``Probabilistic routing in intermittently connected networks,''
  ACM SIGMOBILE Mobile Computing and Communications Review, 
  Vol. 7 Issue 3, pp.19-20(2003).

%MaxProp
\bibitem{Burgess2006}
  J. Burgess, B. Gallagher, D. Jensen, and B. N. Levine, 
  ``MaxProp: Routing for vehicle-based disruption-tolerant networks,''
  In Proc. IEEE INFOCOM, 398-408 (2006).

%OFCOURSE
\bibitem{FM2011b}
  A. Fujihara, S. Ono, and H. Miwa
  ``Optimal Forwarding Criterion of Utility-based Routing under 
  Sequential Encounters for Delay Tolerant Networks,''
  Third International Conference on Intelligent Networking and 
  Collaborative Systems (INCoS) 2011, 279-286 (2011).

\bibitem{FM2013}
  A. Fujihara and H. Miwa, 
  ``Homesick L\'evy Walk and Optimal Forwarding Criterion of Utility-based 
  Routing under Sequential Encounters,'' 
  Internet of things and inter-cooperative computational technologies 
  for collective intelligence, Studies in Computational Intelligence, 
  vol. 460, pp. 207-231 (2013).

\end{thebibliography}

\end{document}